\documentclass{iaus}
\usepackage{graphicx}

\title[Mixing of CNO-cycled matter in active massive stars] 
{Mixing of CNO-cycled matter in pulsationally and magnetically active
massive stars}

\author[Norbert Przybilla \& Maria-Fernanda Nieva]   
{Norbert Przybilla$^1$
 \and Maria-Fernanda Nieva$^2$}

\affiliation{$^1$Dr. Remeis-Observatory \& ECAP, Sternwartstr. 7, 
D-96049 Bamberg, Germany \\
email: {\tt przybilla@sternwarte.uni-erlangen.de} \\[\affilskip]
$^2$Max-Planck-Institut f\"ur Astrophysik, Postfach 1317,
D-85741 Garching, Germany \\ email: {\tt fnieva@mpa-garching.mpg.de}}

\pubyear{2010}
\volume{272}  
\pagerange{1--6}
\jname{Active OB stars -- structure, evolution, mass loss, and critical limits}
\editors{C. Neiner, G. Wade, G. Meynet \& G. Peter, eds.}
\begin{document}

\maketitle

\begin{abstract}
We report on the abundances of helium, carbon, nitrogen and oxygen in
a larger sample of Galactic massive stars of $\sim$7-20\,$M_{\odot}$ near the main
sequence, composed of apparently normal objects, pulsators of $\beta$-Cephei- and
SPB-type, and magnetic stars. High-quality spectra are homogeneously
analysed using sophisticated non-LTE line-formation and comprehensive
analysis strategies. All the stars follow a previously established
tight trend in the N/C-N/O ratio and show normal helium abundances, 
tracing the nuclear path of the
CNO-cycles quantitatively. A correlation of the strength of the mixing
signature with the presence of magnetic fields 
is found. In conjunction with low rotation velocities this
implies that magnetic breaking is highly efficient
for the spin-down of some massive stars. We suggest several objects 
for follow-up spectropolarimetry, as the mixing signature 
indicates a possible magnetic nature of these stars.
\end{abstract}

\firstsection 
\section{Introduction}
Active and normal OB stars occupy the same region in the
Hertzsprung-Russell diagram (HRD). While the observational characteristics
of the different forms of activity are well-defined, a comprehensive
understanding of the physical drivers of activity has to be
established yet. Quantitative analyses of samples of active and normal
stars at high precision are without doubt the starting point for
progress to be made.

Two types of active OB stars are of interest in the following,
{\em pulsators} of $\beta$~Cephei-type and slowly-pulsating B stars (SPBs),
and {\em magnetic stars}. Their surface characteristics will be
compared to those of apparently normal objects. Crucial improvements
over previous work are achieved by the introduction of sophisticated
modelling and a novel analysis technique, which allows in particular
abundance uncertainties to be reduced to about 10-20\% (1-$\sigma$
statistical), in contrast to typical values of a factor $\sim$2 
in literature. 

We focus on surface abundances of C, N and O, which may be altered
already on the main sequence due to mixing with CNO-cycled matter
in the course of the evolution of rotating massive stars
\cite[(e.g. Maeder \& Meynet 2000)]{MM00}. The aim is twofold: {\sc i}) 
to test whether mixing signatures correlate with the presence of 
magnetic fields \cite[(Morel et al. 2008)]{Moreletal08}, and
{\sc ii}) to test whether the occurrence of pulsations is related to
CNO mixing.

\section{Observations and Analysis}
Our observational sample consists of 29 bright 
and apparently slowly rotating early B-type
stars in nearby OB associations and in the field. High-resolution 
($R$\,=\,40-48\,000) and high-S/N spectra ($S/N$\,=\,250-800) 
with wide wavelength coverage 
were obtained with the Echelle spectrographs {\sc Foces}, {\sc Feros}
and {\sc Fies}, and from the {\sc Elodie} archive. Details of the observations 
and the data
reduction are discussed by Nieva \& Przybilla (in prep.) and Nieva,
Sim\'on-D\'iaz \& Przybilla (in prep.).  

The spectra were analysed using a hybrid non-LTE approach
\cite[(Nieva \& Przybilla 2007, 2008; Przybilla et al.
2008b)]{NP07,NP08,PNB08}
and a sophisticated analysis methodology. State-of-the-art atomic input 
data were used in the modelling. In contrast to previous work, 
multiple hydrogen lines, the helium lines, multiple metal ionization
equilibria and the stellar energy distributions were reproduced
{\em simultaneously} in an iterative approach to determine the stellar
atmospheric parameters. Chemical abundances for a wide range of elements 
were derived from analysis of practically the entire
observable spectrum per element. The rewards of such a comprehensive,
but time-consuming procedure were unprecedentedly small statistical
error margins and largely reduced systematics \cite[(Nieva \&
Przybilla~2010)]{NP10}.

\begin{figure}[t]
\begin{center}
\includegraphics[width=11cm]{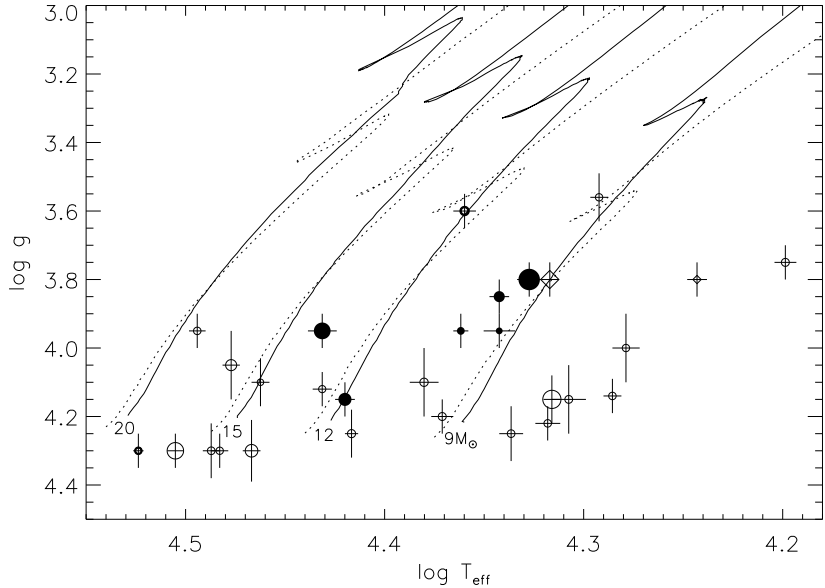}
\caption{The sample stars in the $\log T_\mathrm{eff}-\log g$-plane.
Normal stars are marked by circles, $\beta$\,Cep-type pulsators by
dots (candidates by thick circles) and SPBs by diamonds. The
symbol size encodes the N/C mass ratio, varying between $\sim$0.3
(cosmic value, smallest symbols) and $\sim$1.2. Evolution tracks for
rotating (full lines, $v_\mathrm{ini}$\,=\,300\,km\,s$^{-1}$) 
and non-rotating stars (dotted lines) are overlaid 
\cite[(Meynet \& Maeder~2003)]{MM03}. The tracks were computed
with $Z$\,=\,0.02, while the sample stars have 
$Z$\,=\,0.014, implying a downward shift of the theoretical ZAMS.}
\label{fig1}
\end{center}
\end{figure}

The positions of the sample stars in the $\log T_\mathrm{eff}-\log
g$-plane are displayed in Fig.\,\ref{fig1}. All stars are relatively unevolved
objects on the main sequence in the mass range of $\sim$7-20\,$M_\odot$. 
While the bulk of the objects are normal stars, a small number are known 
pulsators, either of $\beta$~Cephei-type or SPBs. 
Three stars are magnetic, $\beta$\,Cep 
\cite[(Henrichs et al.~2000)]{Henrichsetal00}, $\zeta$\,Cas
\cite[(Neiner et al.~2003)]{Neineretal03} and $\tau$\,Sco 
\cite[(Donati et al.~2006)]{Donatietal06}, one is a candidate magnetic
star, $\delta$\,Cet \cite[(Schnerr et al.~2008; Hubrig et
al.~2008)]{Schnerretal08,Hubrigetal09}. The majority of the stars
has CNO abundances compatible with cosmic values 
\cite[(Przybilla et al.~2008b; Nieva \& Przybilla, in prep.)]{PNB08}, 
about a third of the objects
shows signs of mixing with CNO-cycled matter. We will
concentrate on this aspect and its relation to the presence of
magnetic fields in the following. Fundamental parameters of the sample
stars and also further details on the modelling and analysis
are discussed by Nieva \& Przybilla (these proceedings). 

\section{Signatures of chemical mixing in early B-type stars}
Observational indications of superficial abundance anomalies for
carbon, nitrogen, and oxygen (and the burning product helium) in
OB-type stars on the main sequence are known for a long time.
They have found a theoretical explanation by
evolution models for rotating stars \cite[(e.g. Maeder \& Meynet~2000;
Heger \& Langer~2000)]{MM00,HL00}, where meridional circulation and
turbulent diffusion provide the means to transport CNO-cycled matter
from the stellar core to the surface. The interaction of rotation with
a magnetic dynamo may increase the transport efficiency even further
\cite[(Maeder \& Meynet~2005)]{MM05}.

\begin{figure}[t]
\begin{center}
\includegraphics[width=11cm]{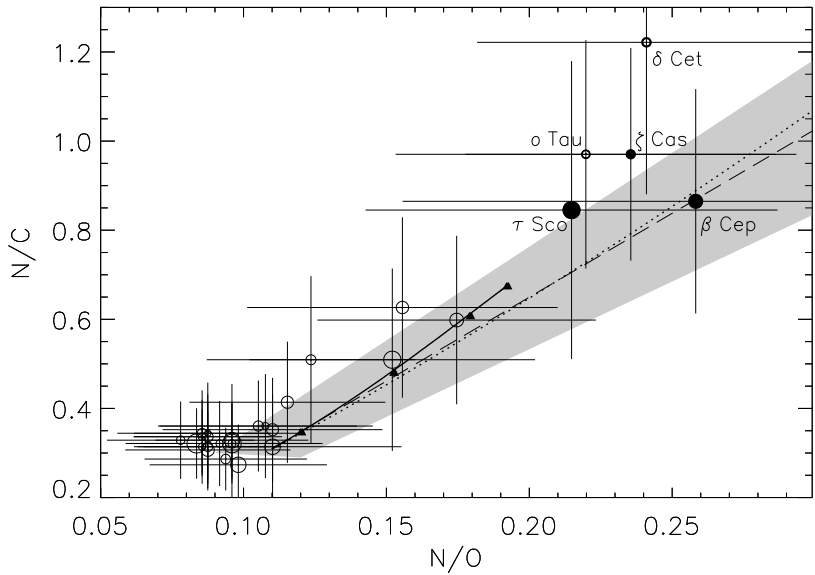}
\caption{N/C vs. N/O abundance ratios (by mass) for our sample stars.
The symbol size encodes the stellar mass, magnetic stars are marked by
filled symbols (candidates by thick circles). Predictions for  
9\,$M_\odot$ models are indicated by the full line, the triangles
marking ratios reached at the end of core-H burning for
$v_\mathrm{ini}$ equivalent to 30, 50, 70 and 90\% of the star's
breakup velocity \cite[(bottom to top, Ekstr\"om et al.~2008)]{Ekstroemetal08}.
The dotted line describes a magnetic 15\,$M_{\odot}$ model of
\cite[Maeder \& Meynet~(2005)]{MM05} with $v_\mathrm{ini}$\,=\,300\,km\,s$^{-1}$.
The long-dashed line corresponds to a slope 3.77 calculated 
analytically for the nuclear path of the CN-cycle and initial CNO
abundances as in the evolution models, while the grey area spans the
full range of theoretical slopes using different references for solar 
abundances and the cosmic abundance standard (upper envelope).}
\label{fig2}
\end{center}
\end{figure}

The changes of CNO abundances relative to each other follow a simple
pattern. At the beginning of hydrogen burning, the $^{14}$N enhancement 
comes from the $^{12}$C destruction via the CN cycle, and the oxygen
content remains about constant. A fraction of this processed material 
is mixed with matter of pristine composition in the course of the
subsequent evolution. A simple calculation shows that the mixing signature 
follows the nuclear path with slope d(N/C)/d(N/O)\,$\approx$\,4 when
abundances are expressed in mass fraction, see \cite{Przybillaetal10}
for details.
Stellar evolution models follow this trend and make quantitative 
predictions concerning the enrichment with nuclear-processed matter.
The mixing signature gets more pronounced with increasing rotational
velocity and increasing mass (not shown here), and magnetic models predict the
strongest mixing, see Fig.~\ref{fig2}. The high-quality analysis of
our sample stars facilitates the predicted trend to be recovered for
the first time \cite[(overplotting data from previous studies would 
fill all areas in the N/O-N/C plane, see Fig.\,1 of Przybilla
et al. 2010, with the error bars exceeding the range of 
Fig.~\ref{fig2} in many cases)]{Przybillaetal10}. About half of the sample clusters around 
(N/O\,$\approx$\,0.09, N/C\,$\approx$\,0.32), which corresponds to unaltered cosmic
values, several more stars show signatures apparently compatible with 
the rotating models, and five stars show a very high degree of mixing.
Of these all but one are known (candidate) magnetic stars, suggesting a  
magnetic nature for the remaining one, $o$\,Tau, as well.
Note that the agreement between observations and theory would be even
better if the models would account for cosmic \cite[($Z$\,=\,0.014,
Przybilla et al. 2008b)]{PNB08} instead of solar abundances
\cite[($Z$\,=\,0.02, Grevesse \& Noels 1993)]{GN93}. 
Helium abundances for all the sample stars are compatible
with cosmic/solar values. 

\begin{figure}[t]
\begin{center}
\includegraphics[width=11cm]{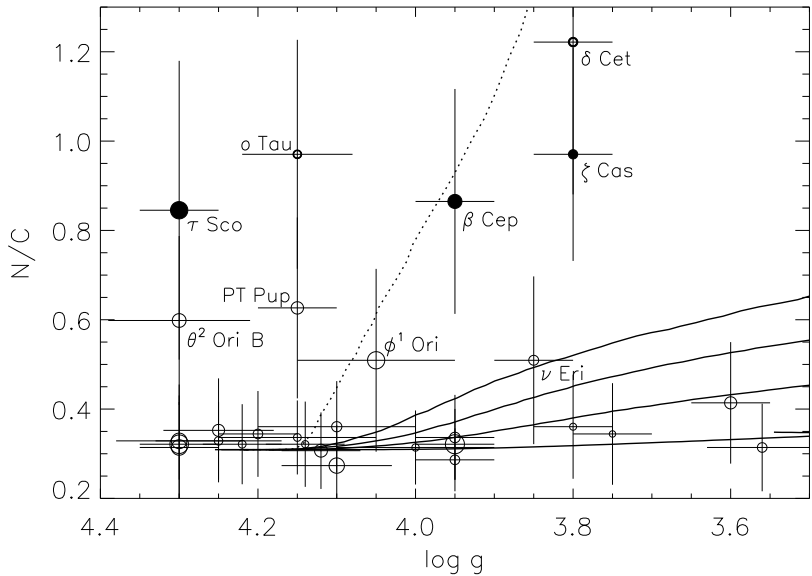}
\caption{N/C abundance ratios on the main sequence as function
of $\log g$. The same symbols as in Fig.~\ref{fig2} are used, the tracks 
for $v_\mathrm{ini}$ of 30, 50, 70 and 90\% breakup velocity (bottom to top)
are separated in this diagram. Several additional stars with their mixing
signature indicating a possible magnetic nature are also identified.
The difference in metallicity between the models ($Z$\,=\,0.02) and 
the sample stars ($Z$\,=\,0.014)
implies a shift of the tracks to the left in this case.}
\label{fig3}
\end{center}
\end{figure}

However, while Fig.\,\ref{fig2} provides a crucial test for the
quality of the analyses it does not facilitate to decide which models 
are more realistic than others, or to identify the mixing mechanism(s). 
Tighter constraints can be put via the $\log g$-N/C diagram (see
Fig.\,\ref{fig3}), which relates the mixing signature to the
evolutionary state of the sample stars. Indeed, the majority of
objects follows the predicted trends for rotating stars, showing only 
mild enrichment (N/C\,$\lesssim$\,0.4) during their evolution from the 
zero-age main sequence (ZAMS, at the highest gravities) to the giant stage 
(lowest gravities). Truely fast rotators (seen pole-on) seem to be
absent, which is likely a relic of one of the selection criteria for the sample 
assembly (stars with low $v\sin i$).

The magnetic stars stand out again with pronounced mixing
signatures (N/C\,$>$\,0.8), complemented by a group of stars at
intermediate values of N/C\,$\approx$\,0.5-0.7.
Some of the signatures found in the more evolved magnetic stars 
like $\beta$\,Cep and $\zeta$\,Cas may be explained
by evolution models accounting for a magnetic dynamo. However, these
models do not predict the magnetic field to reach the stellar surface. 
The magnetic fields are probably of fossil origin instead. It is
likely that the enhanced transport efficiency (when compared to models
accounting for rotation only) is retained in that case, as the basic
physics behind is not sensitive to the origin of the magnetic field.
However, this has to be confirmed by detailed modelling.

Other objects, in particular $\tau$\,Sco, challenge present stellar
models seriously. 
The observed mixing signature of $\tau$\,Sco may be explained by homogeneous
evolution, but this has also to be confirmed by further computations.
Homogeneous evolution would require a highly-efficient spin-down
mechanism, as the star is a truely slow rotator at present 
\cite[(Donati et al.~2006)]{Donatietal06}, like several other magnetic
stars with pronounced mixing signature \cite[(Morel et al. 2008)]{Moreletal08}.
One may speculate on magnetic breaking due to angular-momentum losses by a
magnetically confined line-driven stellar wind or magnetic coupling
to the accretion disc during the star-forming process in the case of a
fossil field. Even though the topic is not understood theoretically in a 
comprehensive way, spin-down times of the order of 1\,Myr 
\cite[(Ud-Doula et al. 2009; Townsend et al.
2010)]{ud-Doula09,Townsend10} or even less
\cite[(Mikul\'a\v{s}ek et al. 2008)]{mikulasek08} 
are reported for some magnetic massive stars,
possibly leading to the required slow rotation already close to the
ZAMS.

After having established the connection of pronounced mixing
signatures with the presence of magnetic fields, one may turn the
argumentation and use the chemical fingerprint as a selection
criterion for the search of magnetic fields in massive stars.
Based on their mixing signatures we therefore suggest the following objects 
for follow-up spectropolarimetric observations: $o$\,Tau as a high-potential 
target and $\theta^2$\,Ori\,B, $\phi^1$\,Ori, PT Pup and $\nu$\,Eri
\cite[(weak constraints for the latter two were presented by
Hubrig et al.~2009)]{Hubrigetal09}. 

Alternative explanations of the
mixing signatures may be required in the case of non-detections. 
Case A mass-transfer in close binary systems -- i.e. near the end of
the core H-burning phase of the primary -- is one of the most
promising mechanisms. This can produce stars highly enriched in
CNO-cycled products close to the ZAMS. Such stars would be fast
rotators because of angular momentum transfer accompanying the mass
overflow. The observational
identification of such systems may be challenging, though, 
as the companion of the visible OB star is likely 
a low-mass helium star in a wide orbit, see
\cite{Wellsteinetal01} for details.\\[-3mm]

Finally, we want to comment briefly on the question whether the occurrence 
of pulsations is in any way related to CNO mixing. The pulsations of
$\beta$\,Cep stars and SPBs are excited via the $\kappa$-mechanism
because of a subsurface opacity bump due to the ionization of 
iron-group elements \cite[(Pamyatnykh 1999)]{Pamyatnykh99}. Yet, an
additional factor is required to explain the occurrence of pulsators
and non-pulastors in the same regions of the HRD. Additional changes
in the opacities due to modified CNO abundances as a consequence of
mixing with nuclear-processed matter may be a possibility. However,
our results do not support this hypothesis, pulsations occur
independently of mixing signatures, see Fig.~\ref{fig1}.

\section{Summary and Outlook}
Theory predicts surface abundances of massive stars to follow a tight 
relation when mixing with CNO-cycled material occurs. The nuclear
path in the N/O-N/C plane is solely determined by the initial CNO
abundances. This provides a powerful criterion for judging the quality of
quantitative analyses. Few early-type massive stars can be 
expected to deviate from this nuclear path, either due to
atomic diffusion (which is normally not operational in OB-type stars) 
or because of pollution with material from advanced nuclear
burning phases, e.g. accreted in the course of a binary supernova
\cite[(Przybilla et al. 2008a; Irrgang et al. 2010)]{PNHB08,Irrgangetal10}.
Only observational data passing this test is well-suited for being used in
the verification of different stellar evolution models.

The high-quality analysis of our sample stars indicates that most
objects can be explained well by existing models for rotating stars.
It also facilitates the identification of a connection of pronounced 
mixing signatures with the presence of magnetic fields --
initially suggested by \cite{Morel08} -- at high confidence.
In reverse, pronounced mixing signatures may be used to define
high-potential targets for spectropolarimetric surveys that aim at the
detection of magnetic stars. Several sample stars are suggested for
follow-up observations.

Further studies, observational as well as theoretical, are 
required to develop a comprehensive understanding of chemical 
mixing in massive stars. High-quality analyses have to be extended to
larger samples of stars, covering a wide range of rotational
velocities. Further contraints to stellar evolution models can be 
obtained by inclusion of stars at different metallicity, like e.g.
objects in the Magellanic Clouds. Once the ongoing surveys as the
MiMeS project (Wade et al., these proceedings) will detect more
magnetic stars it will become feasible to develop a better overview 
on the r\^ole that magnetic dynamos and/or fossil fields play in the
course of the evolution of massive stars.\\[3mm]

\noindent \paragraph{\bf Acknowledgements.} We would like to thank A.
Maeder, G. Meynet, V. Petit, S. Sim\'on-D\'iaz and G. Wade for fruitful
discussion in preparation of this work. NP acknowledges travel support by the
\emph{Deutsche Forschungsgemeinschaft}, project number PR
685/3-1.


\begin{thebibliography}{}

\bibitem[Donati et al. (2006)]{Donatietal06}
{Donati, J.-F., Howarth, I. D., Jardine, M. M., et al.} 2006, \textit{MNRAS}, 370, 629

\bibitem[Ekstr\"om et al. (2008)]{Ekstroem08}
{Ekstr\"om, S., Meynet, G., Maeder, A., \& Barblan, F.} 2008, \textit{A\&A}, 478, 467

\bibitem[Grevesse \& Noels (1993)]{GN93}
{Grevesse, N., \& Noels, A.} 1993, in: S.~Kubono \& T.~Kajino (eds.),
\textit{Origin and Evolution of the Elements}, p.\ 15

\bibitem[Heger \& Langer (2000)]{HL00}
{Heger, A., \& Langer, N.} 2000, \textit{ApJ}, 544, 1016

\bibitem[Henrichs et al. (2000)]{Henrichsetal00}
{Henrichs, H. F., de Jong, J. A., Donati, J.-F., et al.} 2000, \textit{ASP-CS}, 214, 324

\bibitem[Hubrig et al. (2009)]{Hubrigetal09}
{Hubrig, S., Briquet, M., De Cat, P., et al.} 2009, \textit{AN}, 330, 317

\bibitem[Irrgang et al. (2010)]{Irrgangetal10}
{Irrgang A., Przybilla N., Heber U., Nieva M. F., Schuh, S.} 2010, \textit{ApJ}, 711, 138

\bibitem[Maeder \& Meynet (2000)]{MM00}
{Maeder, A., \& Meynet, G.} 2000, \textit{ARAA}, 38, 143
 
\bibitem[Maeder \& Meynet (2005)]{MM05}
{Maeder, A., \& Meynet, G.} 2005, \textit{A\&A}, 440, 1041

\bibitem[Meynet \& Maeder (2003)]{MM03}
{Meynet, G., \& Maeder, A.} 2003, \textit{A\&A}, 404, 975 

\bibitem[Mikul\'a\v{s}ek et al. (2008)]{mikulasek08}
{Mikul\'a\v{s}ek, Z., Krti\v{c}ka, J., Henry, G. W., et al.} 2008, \textit{A\&A}, 485, 585

\bibitem[Morel et al. (2008)]{Morel08}
{Morel, T., Hubrig, S., \& Briquet, M.} 2008, \textit{A\&A}, 481, 453

\bibitem[Neiner et al. (2003)]{Neineretal03}
{Neiner, C., Geers, V. C., Henrichs, H. F., et al.} 2003, \textit{A\&A}, 406, 1019

\bibitem[Nieva \& Przybilla (2007)]{NP07}
{Nieva, M. F., \& Przybilla, N.} 2007, \textit{A\&A}, 467, 295

\bibitem[Nieva \& Przybilla (2008)]{NP08}
{Nieva, M. F., \& Przybilla, N.} 2008, \textit{A\&A}, 481, 199

\bibitem[Nieva \& Przybilla (2010)]{NP10}
{Nieva, M. F., \& Przybilla, N.} 2010, \textit{ASP-CS}, 425, 146

\bibitem[Pamyatnykh (1999)]{Pamyatnykh99}
{Pamyatnykh, A. A.} 1999, \textit{AcA}, 49, 119

\bibitem[Przybilla et al. (2008a)]{PNHB08}
{Przybilla N., Nieva M. F., Heber U., \& Butler K.} 2008a, \textit{ApJ} (Letters), 684, L103 

\bibitem[Przybilla et al. (2008b)]{PNB08}
{Przybilla, N., Nieva, M. F., \& Butler K.} 2008b, \textit{ApJ} (Letters), 688, L103

\bibitem[Przybilla et al. (2010)]{Przybillaetal10}
{Przybilla, N., Firnstein, M., Nieva, M. F., Meynet, G., Maeder, A.} 2010, \textit{A\&A}, 517, A38

\bibitem[Schnerr et al. (2008)]{Schnerretal08}
{Schnerr, R. S., Henrichs, H. F., Neiner, C., et al.} 2008, \textit{A\&A}, 483, 857

\bibitem[Townsend et al. (2010)]{Townsendetal10}
{Townsend, R. H. D., Oksala, M. E., Cohen, D. H., et al.} 2010, \textit{ApJ} (Letters), 714, L318

\bibitem[Ud-Doula et al. (2009)]{ud-Doula09}
{Ud-Doula, A., Owocki, S. P., \& Townsend, R. H. D.} 2009, \textit{MNRAS}, 392, 1022

\bibitem[Wellstein et al. (2001)]{Wellsteinetal01}
{Wellstein, S., Langer, N., Braun, H.} 2001, \textit{A\&A}, 369, 939

\end{thebibliography}
\end{document}